\documentclass[12pt,preprint]{aastex}

\usepackage{amsmath}
\usepackage{amsmath}
\usepackage{amssymb}
\usepackage{amstext}
\usepackage{longtable}
\usepackage{graphicx}
\usepackage{epstopdf}
\usepackage{epsfig}
\usepackage{color}
\definecolor{myColor}{rgb}{0.9,0.9,0.9}  
\begin{document}
\renewcommand\bottomfraction{.9}
\shorttitle{Planetesimal Compositions in Exoplanet Systems} 
\title{Planetesimal Compositions in Exoplanet Systems}

\author{Torrence V. Johnson\altaffilmark{1}, Olivier Mousis\altaffilmark{2,3}, Jonathan I. Lunine\altaffilmark{4}, \& Nikku Madhusudhan\altaffilmark{5}} 
\altaffiltext{1}{Jet Propulsion Laboratory, California Institute of Technology, 4800 Oak Grove Drive, Pasadena, CA, 91109, USA {\tt torrence.v.johnson@jpl.nasa.gov}}
\altaffiltext{2}{Observatoire THETA, Institut UTINAM, UMR 6213 CNRS, Universit{\'e} de Franche-Comt{\'e}, BP 1615, 25010 Besan\c{c}on Cedex, France}
\altaffiltext{3}{Universit\'e de Toulouse; UPS-OMP; CNRS-INSU; IRAP; 14 Avenue Edouard Belin, 31400 Toulouse, France}
\altaffiltext{4}{Center for Radiophysics and Space Research, Space Sciences Building Cornell University,  Ithaca, NY 14853, USA}
\altaffiltext{5}{ Yale Center for Astronomy and Astrophysics, Department of Physics, Yale University, New Haven, CT 06511, USA} 

\begin{abstract}
We have used recent surveys of the composition of exoplanet host stars to investigate the expected composition of condensed material in planetesimals formed beyond the snow line in the circumstellar nebulae of these systems.  Of the major solid forming elements, {C and O abundances (and particularly the C/O abundance ratio) strongly affect the amounts of volatile ices and refractory phases in icy planetesimals formed in these systems.} This results from these elements' effects on the partitioning of O among gas, refractory solid and ice phases in the final condensate. The calculations use a self-consistent model for the condensation sequence of volatile ices from the nebula gas after refractory (silicate and metal) phases have condensed.  The resultant mass fractions (compared to the total condensate) of refractory phases and ices were calculated for a range of nebular temperature structures and redox conditions.   Planetesimals in systems with sub-solar C/O should be water ice-rich, with lower than solar mass fractions of refractory materials, while in super-solar C/O systems planetesimals should have significantly higher mass fractions of refractories, in some cases having little or no water ice.  C-bearing volatile ices and clathrates also become increasingly important with increasing C/O depending on the assumed nebular temperatures.  These compositional variations in early condensates in the outer portions of the nebula will be significant for the equivalent of the Kuiper Belt in these systems, icy satellites of giant planets and the enrichment (over stellar values) of volatiles and heavy elements in giant planet atmospheres.

\end{abstract}


\section{Introduction}

The ``snow line'' is the radial distance in the nebula, variable in time and in height above the midplane, from which (and beyond) water ice becomes thermodynamically stable (Morfill \& V\"{o}lk 1984). Studies of condensation of material from the solar nebula beyond the snow line show that the mass fraction, $f$($s$+$m$), of refractory silicates and metal oxides/sulfides with respect to the total condensate mass depends strongly on the abundances of C and O in the solar nebula, the redox state of C (i.e. CO--rich vs. CH$_4$--rich regions), and the amount of carbon in the form of solid organics {(Wong et al. 2008). The composition of the solid material formed in this region is largely determined by the partitioning of oxygen, the most abundant {solid-forming} element, among gas, ice, and refractory phases.  The amount of available carbon in turn has a strong effect on this partitioning.  The refractory phases considered are {metal oxides and sulfides along with silicates}.  In this context, ``metal'' refers specifically to {all mineral phases containing Fe and Ni, rather than} to all elements heavier than He, the common astrophysical usage.}

In the outer solar nebula, CO is expected to be the dominant C-bearing gas, { whose existence reduces the amount of} O available to form water ice. {In ``warm'' nebula models where mid-plane temperatures remain higher at the end of their evolution (assumed here to be higher than $\sim$50K, see Sec. 4), CO and CH$_4$ remain in the gaseous state and water ice (along with some hydrates) is the primary condensed volatile ice.} In ``cool'' nebula models, i.e. those that are used for modeling the thermodynamic conditions of the solar nebula with mid-plane temperatures { as low as 20 K}, more volatile ices and clathrates, including CO, CH$_4$ and H$_2$S, are added to the water ice and $f$($s$+$m$) is lower (Mousis et al. 2009a). {In circumplanetary environments with higher density and temperature, kinetic considerations are expected to produce more reducing conditions, with CH$_4$ being the primary carbon species (Prinn \& Fegley 1981), resulting in more water ice and lower values of $f$($s$+$m$).}

In extending these concepts on planetary formation and composition to exoplanets, the key issue is stellar composition.  Assuming, as is {commonly done} for the solar system, that the circumstellar disk from which the planets form{ has the same abundance pattern as that of the host star in terms of C/O ratio and overall metallicity (see next paragraph) }, we desire to know the abundances of the major solid-forming elements in the host star.  Stellar type and age provide a general indication of composition, and most host stars in the same galactic environment as the Sun {(excluding high velocity interlopers) } are assumed to have generally ÔsolarÕ or ÔcosmicÕ composition, defined by our best understanding of the composition of the current solar photosphere and proto-solar values (Asplund et al. 2009; Lodders 2003).

The most commonly used figure of merit for stellar compositional differences compared with the Sun is metallicity, or the abundance relative to hydrogen and helium of all elements with atomic number Z$>$2. The logarithmic abundance (dex) of iron relative to hydrogen, [Fe/H], scaled so that the solar value is zero, is commonly used as the working observational definition of metallicity.  In the absence of other compositional information, it is usually assumed that the other heavy elements are present in solar proportion compared to Fe (e.g., Asplund et al. 2009).  Metallicity is known to be positively correlated with the probability of a star having exoplanets, presumably due to the greater amount of planet-building heavier elements (Guillot et al. 2006; Johnson et al. 2010), but it does not provide direct information about the relative proportions of these elements.

However, the detailed compositions of many stars are known to differ significantly from the solar values (Delgado Mena et al. 2010; Petigura \& Marcy 2011; {\"O}berg et al. 2011a; Fortney 2012).  The relative abundance of C and O, {a key determinant of the abundance of water ice in our solar system}, has been discussed as an important factor for planetary systems around other stars, in particular for the composition of Earth-like planets and their potential habitability (Delgado Mena et al. 2010).  For instance, in very carbon-rich systems, carbides may replace silicates and the nebular gas may be water-poor due to the lack of available oxygen (Bond et al. 2010; Gaidos 2000; Lodders 2010; Lodders \& Fegley 1997).{ Spectroscopically determined} stellar compositions for a growing number of stars {with planetary companions} are now becoming available.  In this paper we investigate the role of stellar composition (particularly C and O) on the make-up of planetesimals formed beyond the ÔsnowlineÕ {in these systems}, particularly the relative abundances of refractory solids (silicates and metals) and volatile ices.  In this initial study we have assumed that there is no initial refractory carbon species. We will investigate the effects of solid carbon on condensate composition in subsequent papers.

\section{Stellar Data}

We have used stellar data from four sources with abundance values for the host stars of exoplanet systems and including the most important solid forming elements (C, O, Si, S, Fe and Ni). Source 1 is a survey of 12 stars by Gonzales et al. (2001). Source 2 is a study by Takeda et al. (2001) of 14 stars, including one in common with Source 1 (HD217014/ 51 Peg) and one in common with Source 3 (HD75732/ 55Cnc). Source 3 includes ten stars used for a study by Bond et al. (2010) of diversity in extrasolar terrestrial planets with stellar data (Ecuvillon et al. 2004; Ecuvillon et al. 2006; Gilli et al. 2006).  Finally Source 4 is stellar data for WASP-12 (Fossati et al. 2010), a host star of {a compositionally} interesting transiting planet. Table 1 gives the abundance values for the major solid forming elements from these sources in dex relative to the Sun. Solar abundance values are from Asplund et al. (2009), and given as A(i) = log[n(i)/n(H)] +12 where n(i) is the number of  ``i'' atoms. Table 1 shows the {calculated results for planetesimals forming in a circumstellar nebula with the host starÕs composition for oxidizing (CO--rich) and reducing (CH$_4$--rich) conditions, using the condensation models from Johnson \& Lunine (2005) and Wong et al. (2008).  Values are given for silicate plus metal $f$($s$+$m$) and water ice $f$($Ice$) mass fractions for the simplified case where water ice is the only condensed volatile ice considered. The resulting {material densities for condensates} are also given. As expected, there is a strong correlation between stellar C/O and $f$($s$+$m$) (Johnson et al. 2011).}
  
\section{Models for planetesimal composition}

In the following sections we investigate the composition of planetesimals in exoplanet systems in more detail, including non-water volatile ices and clathrated species and the effects of different assumptions on the circumstellar nebular temperature in the planetesimal formation region.  For this purpose we use the set of stars compiled by Bond et al. (2010) (Source 3 entries in Table 1) covering a wide range of C/O stellar values.  Abundances of Si, Fe, Ni were also reported as well as S for most of the stars.  Solar values of S and P were used if not reported. 
   
Our method combines the calculation of the amount of O taken up in the formation of refractory silicates and metal with a complete thermodynamic treatment describing the formation of ices in protoplanetary disks (Madhusudhan et al. 2011a; Mousis et al. 2009a, 2009b; Mousis et al. 2011).  In our procedure, the O and C remaining in the gas after the formation of silicates and metals were used to calculate the volatile ice condensation chemistry beyond the snow line in the circumstellar nebula.  The result is a set of consistent mass fractions of both refractory (silicate plus metal) and volatile ice mass fractions and the resultant total condensate material density.

\subsection{Refractory phases--silicates and metals}

Oxygen is the dominant solid forming element for refractory solids in the stellar systems studied, forming silicates and ÒmetalÓ (FeO, FeS, Ni), although for very C rich systems, carbides may supplant oxides in the hot inner parts of these systems (Bond et al. 2010; Gaidos 2000).  For our calculations, following Johnson \& Lunine (2005) and Wong et al. (2008), we assume that the silicates are primarily in the form of magnesium silicates, {such as} enstatite (MgSiO$_3$) {or forsterite (Mg$_2$SiO$_4$)}, since Al and Ca are cosmically much less abundant than Mg (which is cosmically similar to Si in abundance).

For each stellar composition we then assume that the number of O atoms in silicates is just three times the Si abundance. {For forsterite, the ratio is four to one, so a mixture of the two will yield an O/Si ratio in the silicate slightly larger but this is not an important effect. } For Fe,  FeS is normally the preferred metal form, but since S is less abundant, any remaining Fe takes up O as FeO, with Ni added to this ÒmetalÓ phase.  This is also supported by the fact that a significant part of S remains as H$_2$S after refractory condensation due to kinetic {effects between the gas and solid phases} (Pasek et al. 2005) in which case somewhat more O would remain as FeO in the refractory phase than in our present calculation.

For each star then, the remaining O in the gas phase {exists in H$_2$O, CO and other volatile molecules}; it is calculated based on the stellar abundances and the {assumed} redox state (see Sec. \ref{Discuss}).  The resulting nebular gas composition is then used as the starting point for calculating the condensation sequence of the volatiles consistent with {the amount of} oxygen tied up in refractory phases.

\subsection{Planetesimal volatile composition}

Our model for the composition of ices formed in various stellar environments is based on a predefined initial gas phase composition in which elemental abundances reflect those of the host star and describes the process by which volatiles are trapped in icy planetesimals formed in the protoplanetary disk. In our model, the oxygen and carbon abundances used are the ones remaining after the formation of refractory phases (silicates and metals) in protoplanetary disks as described above. The process of volatile incorporation in planetesimals, illustrated in Fig. \ref{cool} for various stellar environments, is calculated using the equilibrium curves of hydrates, clathrates and pure condensates, and the ensemble of thermodynamic paths detailing the evolution of temperature and pressure in the 5--20 AU range of the protoplanetary disks. {We refer the reader to Alibert et al. (2005) and Papaloizou \& Terquem (1999) for a full description of the turbulent model of the accretion disk used here. This model {gives the temperature, pressure and density conditions in an evolutionary protoplanetary disk} (dominated by H$_2$) at selected distances to the star as a function of time. The thermodynamic paths represented in Fig. \ref{cool} are then drawn from the temporal evolution of $P$ and $T$ {at a} given distance. The accretion disk model postulates that viscous heating is the predominant heating source, assuming that the outer parts of the disk are protected from solar irradiation by the shadowing effect of { the inner parts of the disk}. Under these conditions, the temperature in the planetesimal-forming region can decrease down to very low values ($\sim$20 K; Mousis et al. 2009a).} For each ice considered in Fig.\ref{cool}, the domain of stability is the region located below its corresponding equilibrium curve. The clathration process stops when no more crystalline water ice is available to trap the volatile species. The equilibrium curves of hydrates and clathrates derive from the compilation of published experimental work contained in (Lunine \& Stevenson 1985), in which data are available at relatively low temperatures and pressures, or a reasonable statistical mechanical model extrapolates the data down to relevant temperatures. On the other hand, the equilibrium curves of pure condensates used in our calculations derive from the compilation of laboratory data given in the CRC Handbook of Chemistry and Physics (Lide 2002). The intersection of a thermodynamic path at a given distance from the star with the equilibrium curves of the different ices allows determination of the amount of volatiles that are condensed or trapped in clathrates at this location in the disk. The volatile, i, to water mass ratio in these planetesimals is determined by the relation (Mousis \& Alibert 2006; Mousis \& Gautier 2004):

\begin{equation}
{m_i = \frac{X_i}{X_{H_2O}} \frac{\Sigma(r; T_i, P_i)}{\Sigma(r; T_{H_2O}, P_{H_2O})}},
\end{equation}

\noindent where $X_i$ and $X_{H_2O}$ are the mass mixing ratios of the volatile $i$ and H$_2$O with respect to H$_2$ in the nebula, respectively. $\Sigma(R; T_i, P_i)$ and $\Sigma(R; T_{H_2O}, P_{H_2O})$ are the surface density of the disk at a distance $r$ from the star at the epoch of hydratation, clathration or condensation of the species $i$, and at the epoch of condensation of water, respectively. From ${\it m_i}$ , it is possible to determine the mass fraction $M_i$ of species $i$ with respect to all the other volatile species taking part to the formation of an icy planetesimal via the following relation:

\begin{equation}
{M_i = \frac{m_i}{\displaystyle \sum_{j=1,n} m_j}},
\end{equation}

\noindent with $\displaystyle \sum_{i=1,n} M_i = 1$.

Note that the ensemble of thermodynamic paths has been arbitrarily chosen to determine the composition of the formed ices. The adoption of any other distance range would not affect the composition of the ices because it remains almost identical irrespective of i) their formation distance and ii) the input parameters of the disk, provided that the initial gas phase composition is homogeneous (Marboeuf et al. 2008).

\section{Discussion}
\label{Discuss}

We have investigated four different cases depicting extreme thermodynamic and chemical conditions in each of our stellar environments. If we assume that viscous heating is the predominant heating source in our disk models and that the outer parts of these disks are protected from solar irradiation by shadowing effect of the inner disk parts, then the temperature in the planet-forming region can decrease down to very low values (typically 20 K, as is assumed for the {solar} nebula -- (Mousis et al. 2009a)). We define these models as the ``cool nebula'' cases. Disk models that include direct stellar radiation and re-radiated stellar energy from the nebula atmosphere have higher midplane temperatures than models dominated by viscous heating. This class of model has been more successful in matching the observed spectral energy distributions of protostellar disks (Dullemond et al. 2007). We define these as the ``warm nebula'' cases where midplane temperature never decrease down below $\sim$50 K. {For the warm nebula model we used the same disk evolution model but simply stopped condensation at the higher midplane temperature.  This is somewhat oversimplified if the goal were to trace composition as a detailed function of time and position within the disk, but does not affect the general results for materials over a broad range of distances beyond the snowline.}

The gas phase composition of the disk can also vary between oxidizing and reducing {states}. In the oxidizing case, we consider a molecular composition of the disk that is similar to that of the {solar} nebula. In this case, oxygen, carbon, nitrogen, sulfur and phosphorus are postulated to exist only in the {molecular species} H$_2$O, CO, CO$_2$, CH$_3$OH, CH$_4$, N$_2$, NH$_3$, H$_2$S and PH$_3$. {We fixed CO/CH$_3$OH/CH$_4$ = 70/2/1 in the gas phase of the disk, a set of values consistent with the ISM measurements made by the Infrared Space Observatory (Ehrenfreund \& Schutte 2000; Gibb et al. 2000) and at millimeter wavelengths from Earth (Frerking et al. 1982; Ohishi et al. 1992) considering the contributions of both gas and solid phases in the lines of sight. On the other hand, the dispersion of the ISM values for the CO/CO$_2$ ratio is in the 0.1--1 range (Gibb et al. 2004; {\"O}berg et al. 2011b) and might reflect object-to-object variation as well as uncertainties of measurements. Here we adopted CO/CO$_2$ = 7/1, {which falls within the range of what has been observed in comets (Bockel{\'e}e-Morvan et al. 2004). Further studies are required to assess the effects of varying this mixing ratio.} We also consider N$_2$/NH$_3$ = 10 in the disk's gas-phase (Lewis \& Prinn 1980). In contrast, in the reducing case, we consider that the remaining C exists only in CH$_4$ and that N$_2$/NH$_3$ = 0.1 in the gas phase. In both cases, { the volatile fraction of} S is assumed to exist in the form of H$_2$S, with H$_2$S/H$_2$ = 0.5 x (S/H$_2$)$_\star$.  This { represents} a maximum H$_2$S mass fraction; if significant S is contained in refractory sulfides, as discussed above, the volatile H$_2$S ice abundance would be proportionately lower.

In the following, we adopt these mixing ratios in our model of the protoplanetary disk. Once the abundances of these molecules have been fixed, the remaining O gives the abundance of H$_2$O. Note that, in the pressure conditions considered in stellar oxidizing environments, CO$_2$ is the only species that crystallizes at a higher temperature than its associated clathrate. We then assume that solid CO$_2$ is the only existing condensed form of CO$_2$ in this environment. In addition, we have considered only the formation of pure ice of CH$_3$OH in our calculations since, to our best knowledge, no experimental data concerning the equilibrium curve of its associated clathrate have been reported in the literature.

\subsection{Planetesimals in a cool nebula ($\sim$20K)}
\subsubsection{Oxidizing conditions}

Figure \ref{cool}a represents the condensation sequence of ices formed down to very low temperatures in oxidizing conditions in the protoplanetary disk surrounding the star HD177830. In this extreme case, the C/O = 0.35 (the lowest value in our stellar population) implies that a substantial amount of water formed in the disk is available for the trapping of volatiles in the form of hydrate or clathrates at intermediary temperatures (in the 50--90 K range). The figure shows that the remaining CO and N$_2$ (not already in clathrate/hydrate) condensate in pure form at very low temperatures. In contrast, Figure \ref{cool}b represents the condensation sequence of ices formed in similar conditions but for C/O = 0.71 in the protoplanetary disk surrounding the star HD108874, i.e. the largest value of our population from which the formation of oxidants is possible. In this other extreme case, only pure condensates form because water is not formed in the disk.

The resultant detailed condensate mass fractions for C/O values between the cases shown in Figs. \ref{cool}a,b, for the case of oxidizing conditions in a cool nebula, are shown in Figure \ref{comp}a.  For subsolar C/O, H$_2$O and silicate plus metal phases dominate, with condensed CO as the next major phase.  With increasing C/O, H$_2$O mass fraction decreases with CO (and lesser amounts of CO$_2$) increasing and becoming the major condensed volatile for super-solar C/O systems.  In addition, condensed N$_2$ is present at mass fractions of a few percent.  Minor species (CH$_3$OH, H$_2$S, NH$_3$, CH$_4$, PH$_3$), with mass fractions of 10$^{-4}$ to 10$^{-2}$, are present in approximately the relative proportions as for the solar nebula.

{ Note that the highest C/O value allowing fully oxidizing conditions is $\sim$0.8 ([C/O]  $\sim$0.16 {dex}). Systems with higher C/O values will be more reducing with CH$_4$ becoming the major C-bearing gas.  As a result the four most C-rich stars in the Bond et al. list (C/O ranging from 0.81 to 1.51) are not included in the plot for oxidizing conditions.}

\subsubsection{Reducing Conditions}
Figure \ref{cool}c represents the condensation sequence of ices formed down to very low temperatures in reducing conditions in the protoplanetary disk surrounding the star HD177830. In this reducing case, for sub-solar C/O = 0.35 (the lowest value in our stellar population), a substantial amount of water formed in the disk is available (somewhat more than in the oxidizing case) for the trapping of volatiles in the form of hydrate or clathrates at { intermediate} temperatures (in the 50--90 K range). Here the only C-bearing volatile is CH$_4$. A substantial part of CH$_4$ is clathrated (about 70\%) and the remaining forms a pure condensate at around 30 K. 

On the other hand, Figure \ref{cool}d represents the condensation of ices formed in the protoplanetary disk surrounding the star HD4203 with C/O = 1.51, the largest value considered in our population. With no significant amount of O in the form of CO, there is sufficient O for some water ice even for very high C/O values, and the figure shows that for reducing conditions the formation conditions of ices { are} similar to { those in} the previous case. NH$_3$ hydrate, H$_2$S, PH$_3$ and CH$_4$ clathrates form at relatively high temperatures until the entire water budget is used and then CH$_4$ and N$_2$ form pure condensates at lower temperatures in the 20--30 K range. 

The resulting mass fractions of condensates for reducing conditions are shown in Figure \ref{comp}b (note the expanded C/O range { for this case, which includes stars with C/O $>$ 0.8)}. H$_2$O and the silicate plus metal are the major phases for sub-solar C/O, with CH$_4$ as clathrate and pure CH$_4$ ice as the other major volatile (becoming comparable to H$_2$O for C-rich systems as more C is available to form CH$_4$) and solid N$_2$ at $<$ 0.01 mass fraction levels.

\subsubsection{Densities}

Condensate densities for the cool nebula cases are shown in Figure \ref{dens}a,b. They are significantly lower than those for the corresponding warm nebula models (see next section) due to the presence of low temperature volatile ices with low densities.   For the solar case, the low densities and large fractions of CO or CH$_4$ ices are generally inconsistent with the properties of current outer solar system bodies such as icy satellites and KBOs.  For example, although Pluto has condensed CO and N$_2$ on its surface (Lellouch et al. 2010), its high density is inconsistent with the 0.25 mass fraction of CO ices in the cool oxidizing nebula case.  Very low temperature condensates such as these may have been present in the early solar nebula and have been suggested as the source of enrichment by icy planetesimals of Jupiter in heavy elements and noble gases (Mousis et al. 2009a). The icy bodies now seen in the outer solar system may have formed under warmer nebular conditions or may have been thermally processed by accretional heating and radiogenic heating in larger planetesimals ($>$ few 10's of km) and lost the bulk of the more volatile ices early on (Castillo-Rogez et al. 2012).

\subsection{Planetesimals in a warm nebula ($>$50K)}
\subsubsection{Oxidizing conditions}

The condensation sequence for volatile ices in the warm nebula follows the same scheme illustrated in Figures \ref{cool}a,b, but terminates at $\sim$50K, so that CO clathrate does not form and CO and N$_2$ remain in the gas phase.

Figure \ref{comp2}a illustrates the relative mass fractions of the condensed phases for oxidizing conditions in a warm nebula over a range of stellar C/O.  For stars with sub-solar C/O values H$_2$O and silicate plus metal dominate the condensate composition with CO$_2$ as the next most abundant species at $<$ $\sim$0.10 mass fraction.  Minor species (CH$_3$OH, H$_2$S, NH$_3$, CH$_4$, PH$_3$), with mass fractions of 10$^{-4}$  to 10$^{-2}$, are present in approximately the relative proportions as for the solar nebula.  As stellar C/O increases, H$_2$O decreases and beyond the solar value ([C/O] =0, C/O = 0.55), rapidly disappears as the C/O $\sim$0.8 is approached, with CO$_2$ and CH$_3$OH ices becoming more important.  Planetesimals in these systems will have refractory, silicate plus metal rich compositions compared with solar system conditions.  

\subsubsection{Reducing conditions}

In the C-rich, ``oxygen-starved'' circumstellar nebulae, conditions will shift to more reducing conditions, kinetic considerations permitting, and H$_2$O will remain the primary condensed volatile, with CH$_4$ and NH$_3$ (in clathrate and hydrated phases) replacing CO$_2$ and CH$_3$OH as secondary condensed volatiles as shown in Figures \ref{cool}c,d.  Solid C phases may also be present in C-rich systems, although we have not explored these cases in this paper.  Figure \ref{comp2}b shows the mass fractions for the fully reducing case (applicable also to condensates in circumplanetary sub-nebulae in these systems -- see Prinn \& Fegley 1989).  It is interesting to note that the mass fraction of CH$_4$ in planetesimals actually decreases with increasing C/O under these conditions.  This is due to the fact that in the warm nebula CH$_4$ is incorporated only as clathrate in the water ice structure and as the amount of H$_2$O decreases in O depleted systems, so also does CH$_4$ in the condensates, with more remaining in the gas phase.

\subsubsection{Densities}

Figure \ref{dens2}a,b shows the material density of planetesimal condensates for the warm nebula cases. { In contrast to the {cool-nebula} case (see Sec. 4.1.3) } the densities for the solar oxidizing case ($\sim$2200 kg/m$^3$) are similar to the uncompressed densities of outer solar system icy bodies such as Pluto, Triton and Phoebe,  and the solar reducing densities ($\sim$1400 kg/m$^3$)  are similar to the uncompressed densities of large icy satellites that may have formed in {reducing conditions in circumplanetary disks} (Johnson \& Estrada 2009; Johnson \& Lunine 2005; Wong et al. 2008). {This is consistent with the suggestion in section 4.1.3 that these bodies formed from material condensed under warm nebular conditions or subsequently lost most low temperature condensates.}

\subsection{Effects of other stellar elemental abundances}

{As mentioned in the introduction, we expect the C and O abundances to have their greatest effect on the overall refractory and volatile ice composition of planetesimals through limiting the O available for different solid phases during condensation.  Other abundances may affect the detailed composition for minor species but should not have a systematic effect on the major phases.  This is illustrated in Figures \ref{cool_oxy}, \ref{cool_red}, \ref{warm_oxy} and \ref{warm_red} where data from Figures \ref{comp} and \ref{comp2} for species with mass fraction $>$ 0.01 are plotted both against [C/O] and [Fe/H].  The [Fe/H] plots show a low degree of correlation with the mass fractions of major species in contrast to the strong correlations with [C/O].  This is true for both oxidizing/reducing and warm/cool cases.  The only apparent exception is for the minor N-bearing species N$_2$ and NH$_3$ where the N phase mass fractions decrease with increasing stellar [Fe/H]. This apparent correlation is an artifact of the lack of measured host-star N abundances. {Since stellar N abundance values are not available, this is simply the result of assuming a fixed solar [N/H] value combined with different [Fe/H] for each star, so the ratio [N/Fe] decreases for increasing  Fe abundance.} %Solar values for N were used {to estimate the general  phase behavior of N, and increasing Fe abundances resulted in less N in the condensation calculations.}}

\section{Wasp12b}

Wasp12b has been shown to have a very C-rich atmosphere, with C/O $\sim$1 (Madhusudhan et al. 2011b).  The host star, Wasp12, is however unremarkable in its slightly sub-solar C/O ratio, with [C/O] = -0.09 dex.  Madhusudhan et al. (2011a) have shown that Wasp12b's atmospheric C/O is inconsistent with an atmosphere of the host star's composition even if enriched in heavy elements by icy planetesimals dissolved in the planet's {envelope} and suggest that it may have been formed in {an} O-depleted part of the circumstellar nebula.  The self-consistent calculations of silicate, metal and volatile ice fractions in this paper can be used to extend the conclusions of Madhusudhan et al. (2011a). Figure \ref{comp}a,b and 4a,b show the composition of planetesimals in a nebula with the stellar composition of Wasp12 (light grey vertical line) for temperature and redox conditions discussed above.  Using these relative proportions of heavy elements, the basic conclusions remain unchanged since the enriching planetesimals still retain the C/O signature of the stellar atmosphere unless there is an oxygen-depleted region in the planet-forming zone as suggested in Madhusudhan et al. (2011a).

\section{Summary}

Planetary systems around stars with {compositions that differ} from the {Sun's} will have planetesimals -- formed beyond the ÔÕsnow lineÕÕ-- which have a wide range of silicate and metal, carbon and ice proportions. The fraction of silicate plus metal in extrasolar planetesimals should depend strongly on the C/O ratio in {a given} circumstellar nebula, which controls the abundance of water ice, clathrate hydrates, and stochiometric hydrates in the condensed solids. Other volatile ices are less strongly affected by the host star compositions in our study, although C-bearing ices such as CO (ice and clathrate), CO$_2$ and CH$_3$OH will be more abundant with increasing C/O. The planetesimal population of our own outer solar system, as incompletely known as it is, represents only one {trajectory} through the planetesimal composition space defined by the possible range of C/O and metallicities seen in other stars.  

These characteristics of exoplanet condensates may be investigated in the future through their {effect} on heavy element enrichment and hence atmospheric compositions of extrasolar giant planets (e.g., Mousis et al. 2009a; Mousis et al. 2011) and possibly through {compositional} data for material in planet-forming zones in young systems. Existing telescopic systems can provide constraints on such compositions, but the James Webb Space Telescope will have much greater sensitivity and resolution allowing for a much larger number of giant planet and disk compositions {to be measured}. 

\acknowledgements{TVJ work was done at the Jet Propulsion Laboratory, California Institute of Technology under a contract from NASA.  Government sponsorship acknowledged. NM acknowledges support from NASA HST and JPL/Spitzer grants. JIL was supported by the James Webb Space Telescope Project through NASA. O.M. acknowledges support from CNES. We also thank an anonymous referee for helpful comments and Dr. Neil Turner for discussions of the warm nebula models.}

\clearpage
\begin{table}
\caption[]{Star Compositions for Exoplanet Hosts (data from different sources -- see text)}
\begin{center}
\rotatebox{0}{
\scalebox{0.4}{
\begin{tabular}{lcccccccccc}
\hline
\hline
\noalign{\smallskip}
Source		& 1			& 1			& 1			& 1			& 1			& 1			& 1			& 1			& 1			& 1				\\					
			& HD 10697	& HD 12661	& HD 52265	& HD 89744	& HD 130322	& HD 134987	& HD 168443	& HD 92263	& HD 209458	& HD 217014 51 Peg	\\
\noalign{\smallskip}
\hline
\noalign{\smallskip}
C			& 0.18		& 0.35		& 0.12		& 0.16		& 0.02		& 0.34		& 0.23		& 0.42		& -0.09		& 0.21			\\
O			& 0.27		& 0.27		& 0.2			& 0.24		& 0.1			& 0.38		& 0.22		& 0.3			& 0.1			& 0.26			\\
Si			& 0.15		& 0.38		& 0.26		& 0.3			& 0.03		& 0.32		& 0.15		& 0			& 0.05		& 0.22			\\
S			& 0.19		& 0.39		& 0.11		& 0.08		& 0.18		& 0.34		& 0.21		& 0.05		& -0.12		& 0.18			\\
Fe			& 0.09		& 0.28		& 0.19		& 0.23		& -0.02		& 0.25		& 0.03		& -0.02		& -0.03		& 0.14			\\
Ni			& 0.12		& 0.33		& 0.21		& 0.24		& 0.01		& 0.32		& 0.13		& -0.2		& 0			& 0.19			\\
C/O			& -0.09		& 0.08		& -0.08		& -0.08		& -0.08		& -0.04		& 0.01		& 0.12		& -0.19		& -0.05			\\
C/Fe			& 0.09		& 0.07		& -0.07		& -0.07		& 0.04		& 0.09		& 0.2			& 0.44		& -0.06		& 0.07			\\
Oxidizing		&			&			&			&			&			&			&			&			&			& 				\\					
$f$($s$+$m$)	&0.56		& 0.93		& 0.73		& 0.73		& 0.61		& 0.65		& 0.67		& 0.67		& 0.56		& 0.65			\\
$f$($Ice$)		&0.44		& 0.07		& 0.27		& 0.27		& 0.39		& 0.35		& 0.33		& 0.33		& 0.44		& 0.35			\\
$\rho$		&1.63		& 3.18		& 2.09		& 2.09		& 1.74		& 1.84		& 1.90		& 1.92		& 1.64		& 1.86			\\
Reducing		&			&			&			&			&			&			&			&			&			&				\\						
$f$($s$+$m$)	&0.37		& 0.53		& 0.50		& 0.50		& 0.41		& 0.41		& 0.39		& 0.27		& 0.42		& 0.42			\\
$f$($Ice$)		&0.63		& 0.47		& 0.50		& 0.50		& 0.59		& 0.59		& 0.61		& 0.73		& 0.58		& 0.58			\\
$\rho$		&1.31		& 1.56		& 1.51		& 1.52		& 1.36		& 1.36		& 1.33		& 1.19		& 1.37		& 1.38			\\
\noalign{\smallskip}
\hline
\noalign{\smallskip}
Source		& 1			& 1			& 2			& 2			& 2			& 2			& 2			& 2			& 2			& 2 				\\
			&HD 217107	&HD 222582	&Tau Boo		&$\nu$ And	& 16 Cyg B	& Epsilon Eri	& 47 UMa		& HD 89744	& HD 217014 51 Peg	& HD 75732 55 Cnc			\\
\noalign{\smallskip}
\hline
\noalign{\smallskip}
C			& 0.31		& -0.07		& 0.46		& 0.33		& 0.19		& 0.25		& 0.19		& 0.31		& 0.37				& 0.22			\\
O			& 0.29		& 0.12		& 0.28		& 0.21		& 0.21		& 0.14		& 0.2			& 0.2			& 0.34				& 0.46			\\
Si			& 0.34		& 0.07		& 0.18		& 0.1			& 0.01		& -0.12		& -0.07		& 0.18		& 0.08				& 0.18			\\
S			& 0.41		& 0.06		& 0.56		& 0.34		& 0.17		& 0.29		& 0.18		& 0.31		& 0.38				& 0.63			\\
Fe			& 0.29		& -0.05		& 0.18		& 0.1			& 0.05		& -0.1		& -0.08		& 0.21		& 0.19				& 0.27			\\
Ni			& 0.32		& -0.03		& 0.33		& 0.21		& 0.19		& -0.04		& 0.05		& 0.25		& 0.38				& 0.59			\\
C/O			& 0.02		& -0.19		& 0.18		& 0.12		& -0.02		& 0.11		& -0.01		& 0.11		& 0.03				& -0.24			\\
C/Fe			& 0.02		& -0.02		& 0.28		& 0.23		& 0.14		& 0.35		& 0.27		& 0.1			& 0.18				& -0.05			\\
Oxidizing		&			&			&			&			&			&			&			&			&					& 				\\					
$f$($s$+$m$)	& 0.81		& 0.55		& 0.97		& 0.83		& 0.58		& 0.68		& 0.51		& 0.90		& 0.60				& 0.42			\\
$f$($Ice$)		& 0.19		& 0.45		& 0.03		& 0.17		& 0.42		& 0.32		& 0.49		& 0.10		& 0.40				& 0.58			\\
$\rho$		& 2.41		& 1.60		& 3.64		& 2.55		& 1.69		& 1.94		& 1.54		& 3.00		& 1.75				& 1.38			\\
Reducing		&			&			&			&			&			&			&			&			&					&				\\						
$f$($s$+$m$)	& 0.49		& 0.40		& 0.40		& 0.39		& 0.35		& 0.30		& 0.29		& 0.47		& 0.33				& 0.32			\\
$f$($Ice$)		& 0.51		& 0.60		& 0.60		& 0.61		& 0.65		& 0.70		& 0.71		& 0.53		& 0.67				& 0.68			\\
$\rho$		& 1.50		& 1.35		& 1.35		& 1.34		& 1.28		& 1.22		& 1.21		& 1.46		& 1.26				& 1.24			\\
\noalign{\smallskip}
\hline
\noalign{\smallskip}
Source		& 2			& 2			& 2			& 2				& 2			& 2			& 3					& 3			& 3			& 3				\\
			& 70 Vir		& Rho CrB	& HD 217107	& HD 52265		& HD 38529	& 14 Her		& HD 75732 55Cnc		& Gl777A		& HD 4203	& HD 17051		\\
\noalign{\smallskip}
\hline
\noalign{\smallskip}
C			& -0.04		& -0.1		& 0.47		& 0.32			& 0.4			& 0.57		& 0.28				& 0.26		& 0.42		& 0.36			\\
O			& 0.1			& 0.12		& 0.43		& 0.28			& 0.37		& 0.49		& 0.11				& 0.2			& -0.02		& 0.09			\\
Si			& -0.09		& -0.21		& 0.19		& 0.16			& 0.27		& 0.28		& 0.31				& 0.26		& 0.46		& 0.25			\\
S			& 0.17		& 0.05		& 0.64		& 0.37			& 0.63		& 0.91		& 0.15				& 0.13		& 0.23		& -0.02			\\
Fe			& -0.08		& -0.3		& 0.28		& 0.19			& 0.37		& 0.39		& 0.29				& 0.2			& 0.36		& 0.2				\\
Ni			& 0			& -0.21		& 0.5			& 0.34			& 0.54		& 0.68		& 0.33				& 0.27		& 0.44		& 0.29			\\
C/O			& -0.14		& -0.22		& 0.04		& 0.04			& 0.03		& 0.08		& 0.17				& 0.06		& 0.44		& 0.27			\\
C/Fe			& 0.04		& 0.2			& 0.19		& 0.13			& 0.03		& 0.18		& -0.01				& 0.06		& 0.06		& 0.16			\\
Oxidizing		&			&			&			&				&			&			&					&			&			& 				\\					
$f$($s$+$m$)	& 0.49		& 0.34		& 0.61		& 0.70			& 0.72		& 0.68		& 1.21				& 0.88		& 2.77		& 1.62			\\
$f$($Ice$)		& 0.51		& 0.66		& 0.39		& 0.30			& 0.28		& 0.32		& 0.00				& 0.12		& 0.00		& 0.00			\\
$\rho$		& 1.51		& 1.26		& 1.77		& 2.03			& 2.10		& 1.95		& 3.23				& 2.80		& 1.39		& 2.40			\\
Reducing		&			&			&			&				&			&			&					&			&			&				\\						
$f$($s$+$m$)	& 0.34		& 0.25		& 0.34		& 0.40			& 0.43		& 0.36		& 0.63				& 0.50		& 0.86		& 0.59			\\
$f$($Ice$)		& 0.66		& 0.75		& 0.66		& 0.60			& 0.57		& 0.64		& 0.37				& 0.50		& 0.14		& 0.41			\\
$\rho$		& 1.27		& 1.15		& 1.27		& 1.35			& 1.40		& 1.29		& 1.81				& 1.52		& 2.69	 	& 1.70			\\
\noalign{\smallskip}
\hline
\noalign{\smallskip}
Source		& 3			& 3			& 3			& 3				& 3			& 3			& 4			& 			&			& 			\\
			& HD 19994	& HD 27442	& HD 72659	& HD 08874		& HD 177830	& HD 213240	& WASP 12	& Solar abundance (Asplund et al. 2009)		&			& 			\\
\noalign{\smallskip}
\hline
\noalign{\smallskip}
C			& -0.14		& 0.18		& 0.45		& 0.25		& 0.14		& 0.31		& 0.08			& 8.47		&			& 			\\
O			& 0.09		& -0.12		& 0.36		& 0.14		& 0.33		& 0.34		& 0.17			& 8.73		&			& 			\\
Si			& 0.07		& 0.16		& 0.33		& 0.21		& 0.11		& 0.49		& 0.05			& 7.54		&			& 			\\
S			& -0.2		& 0.03		& 0.03		& 0.03		& -0.07		& 0.03		& 0.18			& 7.16		&			& 			\\
Fe			& -0.01		& 0.19		& 0.29		& 0.22		& 0.13		& 0.35		& 0.43			& 7.54		&			& 			\\
Ni			& 0.03		& 0.2			& 0.41		& 0.21		& 0.15		& 0.38		& 0.27			& 6.26		&			& 			\\
C/O			& -0.23		& 0.3			& 0.09		& 0.11		& -0.19		& -0.03		& -0.09			&			&			& 			\\
C/Fe			& -0.13		& -0.01		& 0.16		& 0.03		& 0.01		& -0.04		& -0.35			&			&			& 			\\
Oxidizing		&			&			&			&			&			&			&				&			&			& 			\\							
$f$($s$+$m$)	& 0.57		& 1.70		& 0.86		& 0.99		& 0.46		& 0.84		& 0.73			&			&			& 			\\
$f$($Ice$)		& 0.43		& 0.00		& 0.14		& 0.01		& 0.54		& 0.16		& 0.27			&			&			& 			\\
$\rho$		& 1.66		& 2.33		& 2.71		& 3.82		& 1.45		& 2.57		& 2.19			&			&			& 			\\
Reducing		&			&			&			&			&			&			&				&			&			& 			\\					
$f$($s$+$m$)	& 0.44		& 0.72		& 0.44		& 0.53		& 0.33		& 0.56		& 0.53			&			&			& 			\\
$f$($Ice$)		& 0.56		& 0.28		& 0.56		& 0.47		& 0.67		& 0.44		& 0.47			&			&			& 			\\
$\rho$		& 1.41		& 2.07		& 1.42		& 1.58		& 1.26		& 1.62		& 1.60			&			&			& 			\\
\noalign{\smallskip}
\hline
\end{tabular}}}
\end{center}
\label{data}
\end{table}

\clearpage
\begin{figure}
\centering
\resizebox{\hsize}{!}{\includegraphics[angle=90]{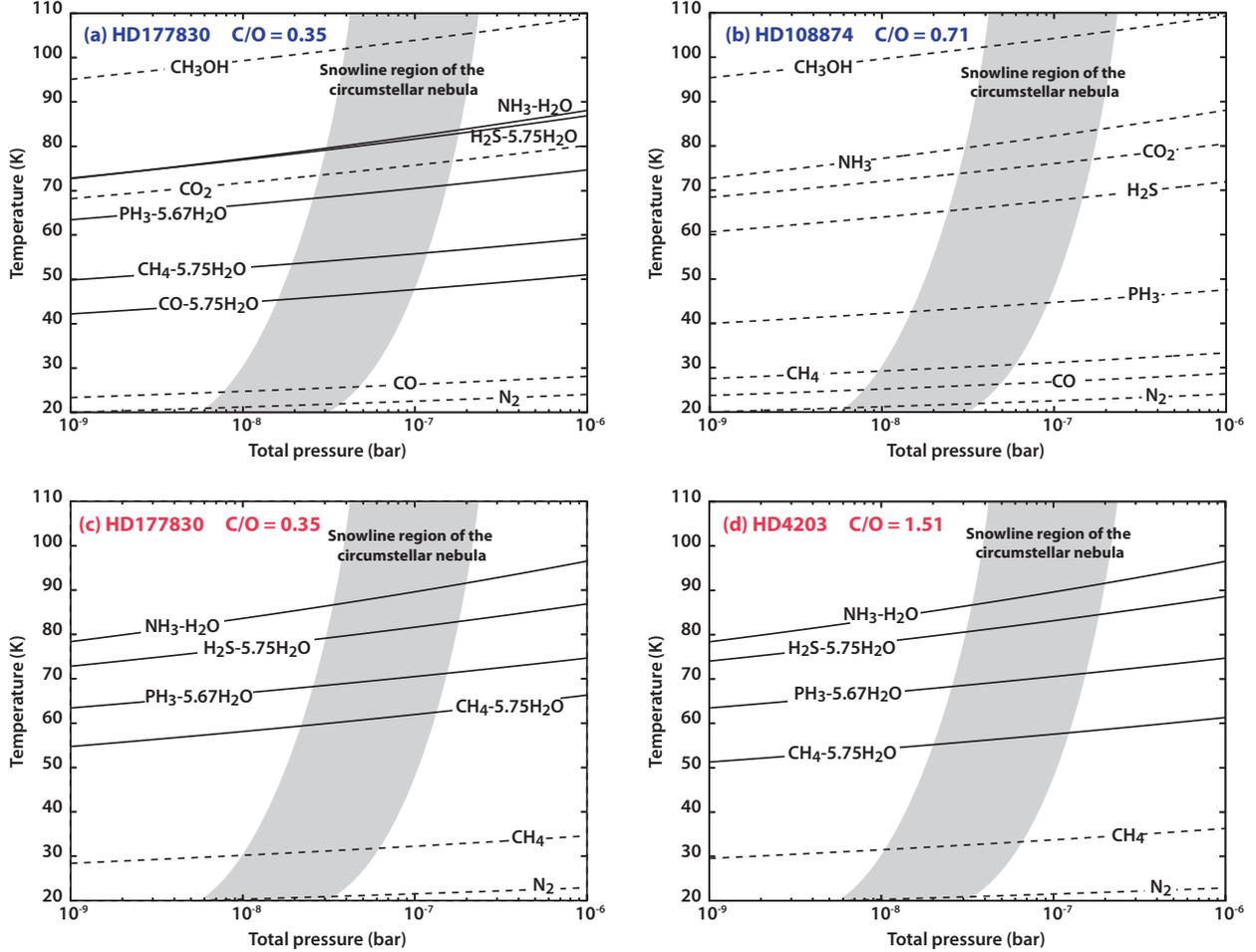}}
\caption{Condensation sequence for volatile ices in planetesimals formed in the snowline region of the protoplanetary disks surrounding exoplanet host stars, assuming a full efficiency of clathration. Species remain in the gas phase above the equilibrium curves. Below, they are trapped as clathrates (X-5.75H$_2$O or X-5.67H$_2$O) or form hydrate (NH$_3$-H$_2$O) (solid lines), or simply condense (dotted lines) when water is not anymore available. The stellar examples shown, their C/O value, and the assumed redox conditions are: a) HD177830 (C/O = 0.35), Oxidizing.  b) HD108874 (C/O = 0.71), Oxidizing. In this case, water does not exist in the disk and only pure condensates form (dotted lines).  c) HD177830 (C/O = 0.35), Reducing. d) HD4203 (C/O = 1.51), Reducing.}
\label{cool}
\end{figure}

\clearpage
\begin{figure}
\centering
\includegraphics[angle=0,width=11cm]{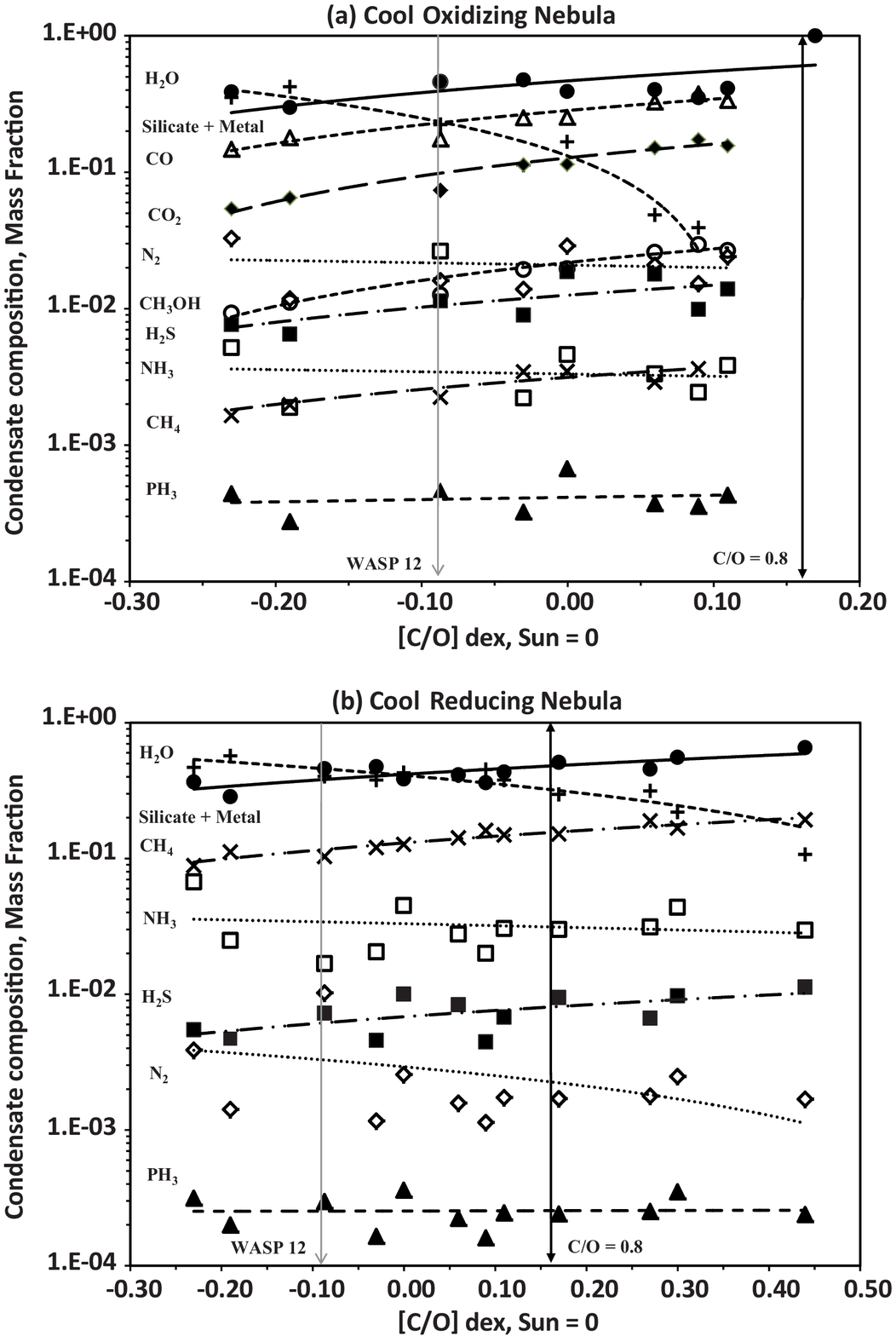}
\caption{Planetesimal compositions for cool ($<$ 20K) nebula case: Condensate relative mass fractions of refractory phases (silicate plus metal) and volatile ices in planetesimals formed in circumstellar nebulae around host stars with varying C/O values, expressed as in logarithmic units [C/O] dex with the Solar value = 0.  Dark vertical line marks value for C/O = 0.8 number ratio, approximately the point at which all the oxygen is taken up by silicates, metals, and CO and more C-rich systems must become more reducing.  Light vertical line marks the C/O value for Wasp12, a host star with a carbon-rich planet in spite of the sub-solar stellar C/O abundance. a) Oxidizing conditions. b) Reducing conditions  as described in the text.}
\label{comp}
\end{figure}

\clearpage
\begin{figure}
\centering
\includegraphics[angle=0,width=12cm]{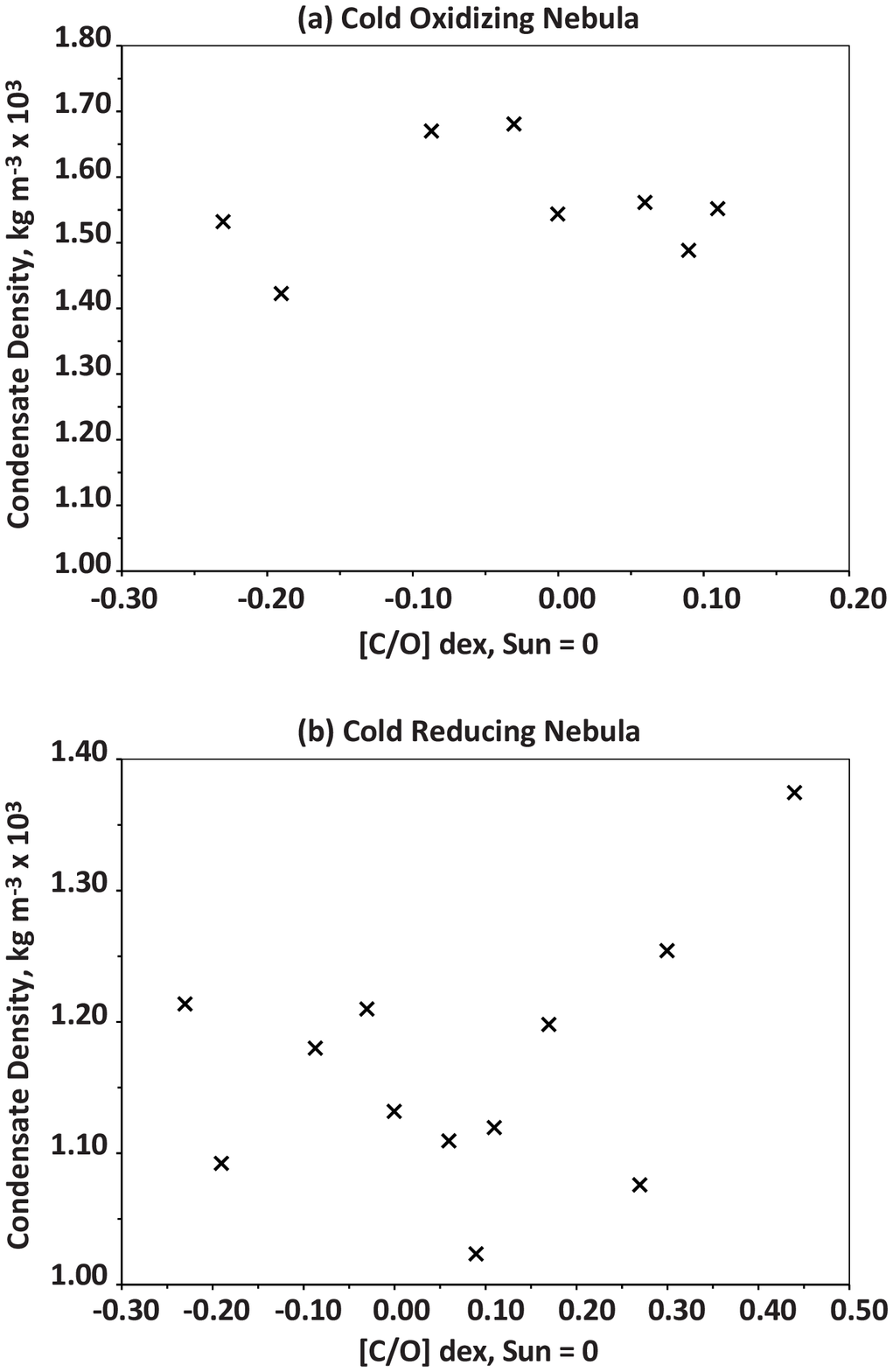}
\caption{Material density of condensates for cool ($<$20K) nebula case:  Density in kg/m$^3$ $\times$ 10$^3$ for condensates (no porosity) formed in circumstellar nebulae around host stars with varying C/O values, expressed as in logarithmic units [C/O] dex with the Solar value = 0. a) Oxidizing conditions. b) Reducing conditions.}
\label{dens}
\end{figure}

\clearpage
\begin{figure}
\centering
\includegraphics[angle=0,width=13cm]{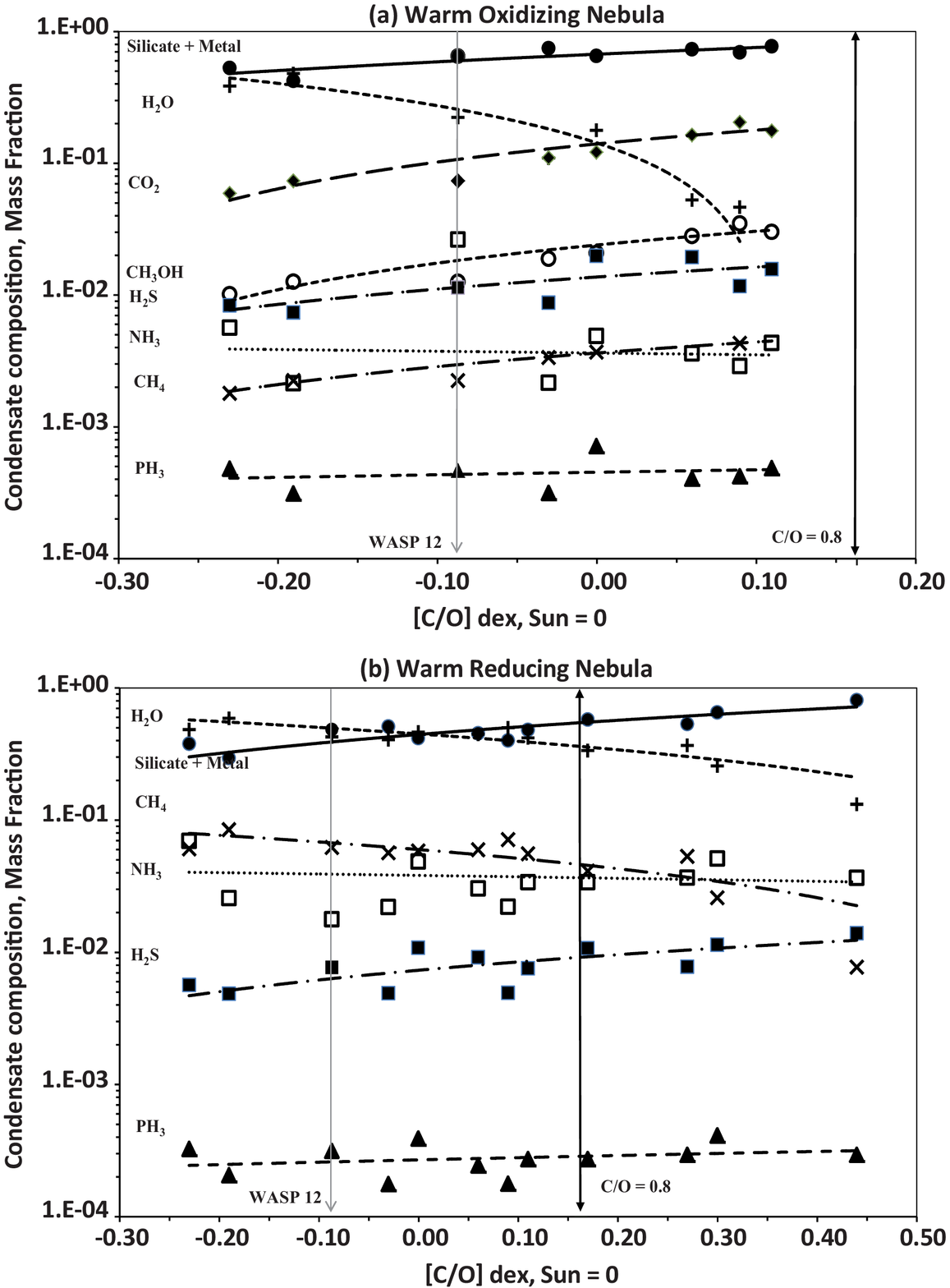}
\caption{Planetesimal compositions for warm ($\sim$50K) nebula case: Condensate relative mass fractions of refractory phases (silicate plus metal) and volatile ices in planetesimals formed in circumstellar nebulae around host stars with varying C/O values, expressed as in logarithmic units [C/O] dex with the Solar value = 0.  Dark vertical line marks value for C/O = 0.8 number ratio, approximately the point at which all the oxygen is taken up by silicates, metals, and CO and more C-rich systems must become more reducing. Light vertical line marks the C/O value for Wasp12, a host star with a carbon-rich planet in spite of the sub-solar stellar C/O abundance. a) Oxidizing conditions. b) Reducing conditions  as described in the text.}
\label{comp2}
\end{figure}

\clearpage
\begin{figure}
\centering
\includegraphics[angle=0,width=13cm]{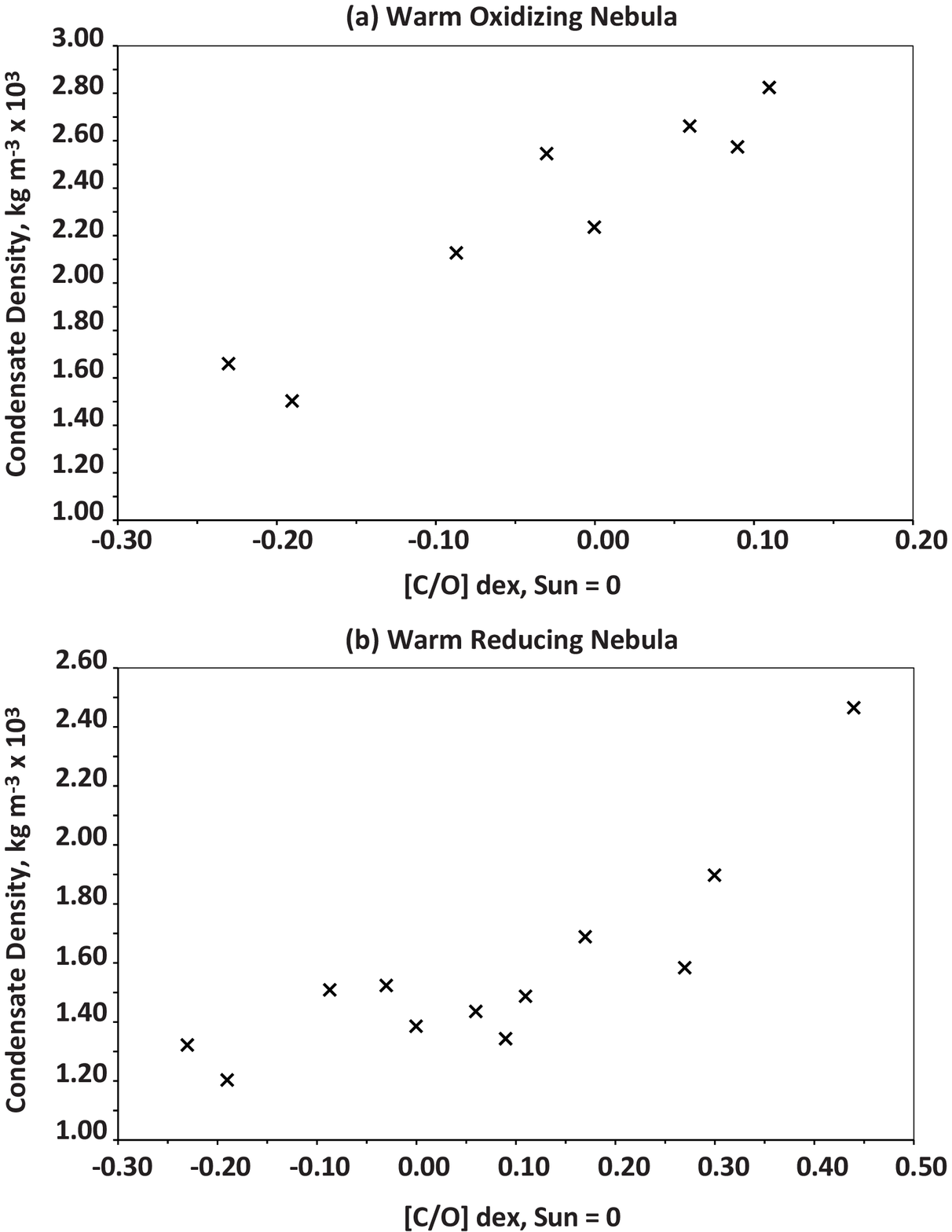}
\caption{Material density of condensates for warm ($\sim$50K) nebula case:  Density in kg/m$^3$ $\times$ 10$^3$ for condensates (no porosity) formed in circumstellar nebulae around host stars with varying C/O values, expressed as in logarithmic units [C/O] dex with the Solar value = 0. a) Oxidizing conditions. b) Reducing conditions.}
\label{dens2}
\end{figure}

\clearpage
\begin{figure}
\centering
\includegraphics[angle=0,width=13cm]{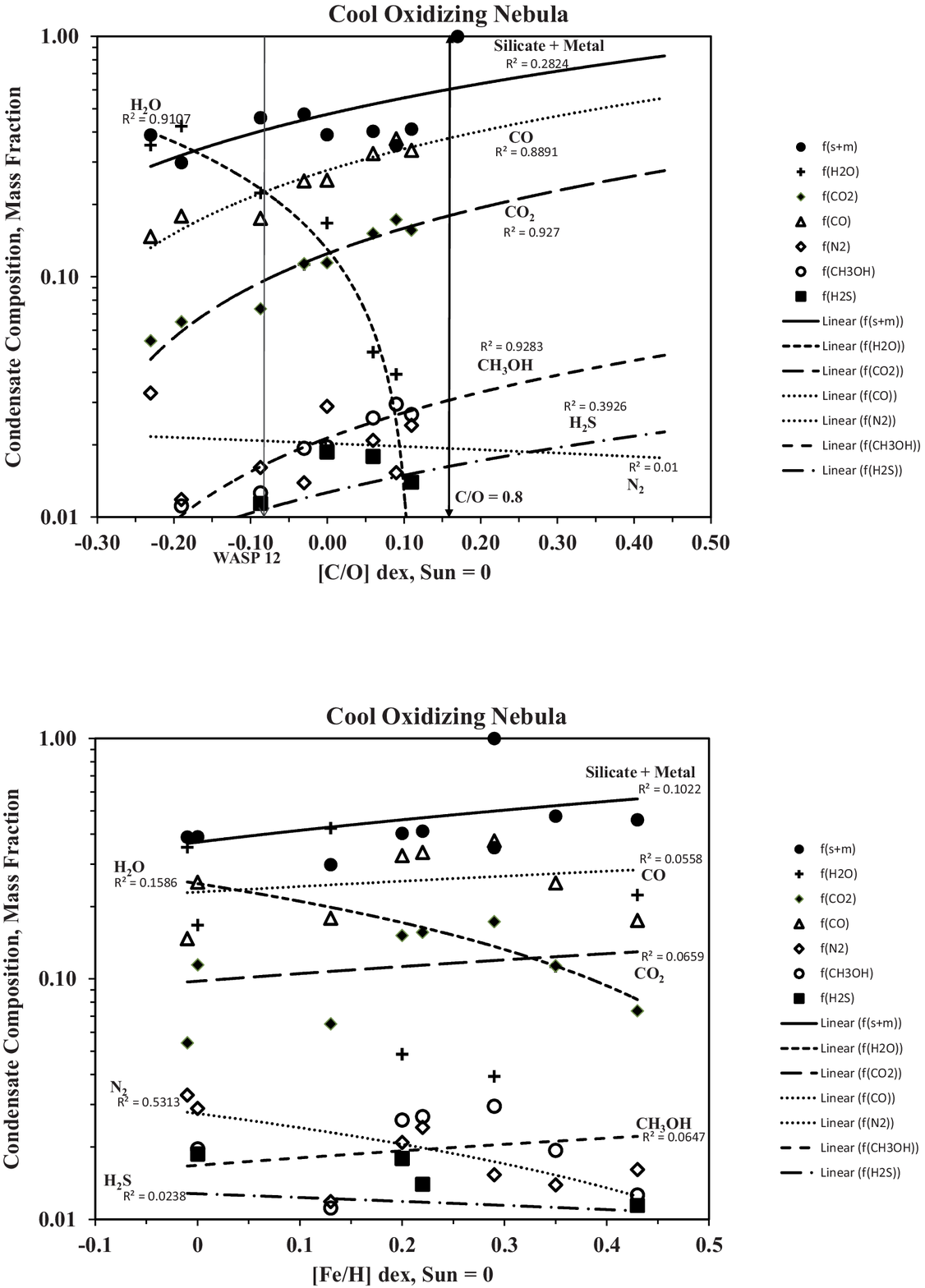}
\caption{Cool nebula case and oxidizing conditions.  Correlation of mass fractions of major ($>$ 0.01 on y-axis) refractory and volatile ice phases with [C/O] and [Fe/H].  Data are the same as Fig. \ref{comp}a,b.  The correlation with [C/O] is much stronger than with [Fe/H] as indicated by the scatter in the calculated values and the R$^2$ values for the trendline analysis shown.}
\label{cool_oxy}
\end{figure}

\clearpage
\begin{figure}
\centering
\includegraphics[angle=0,width=13cm]{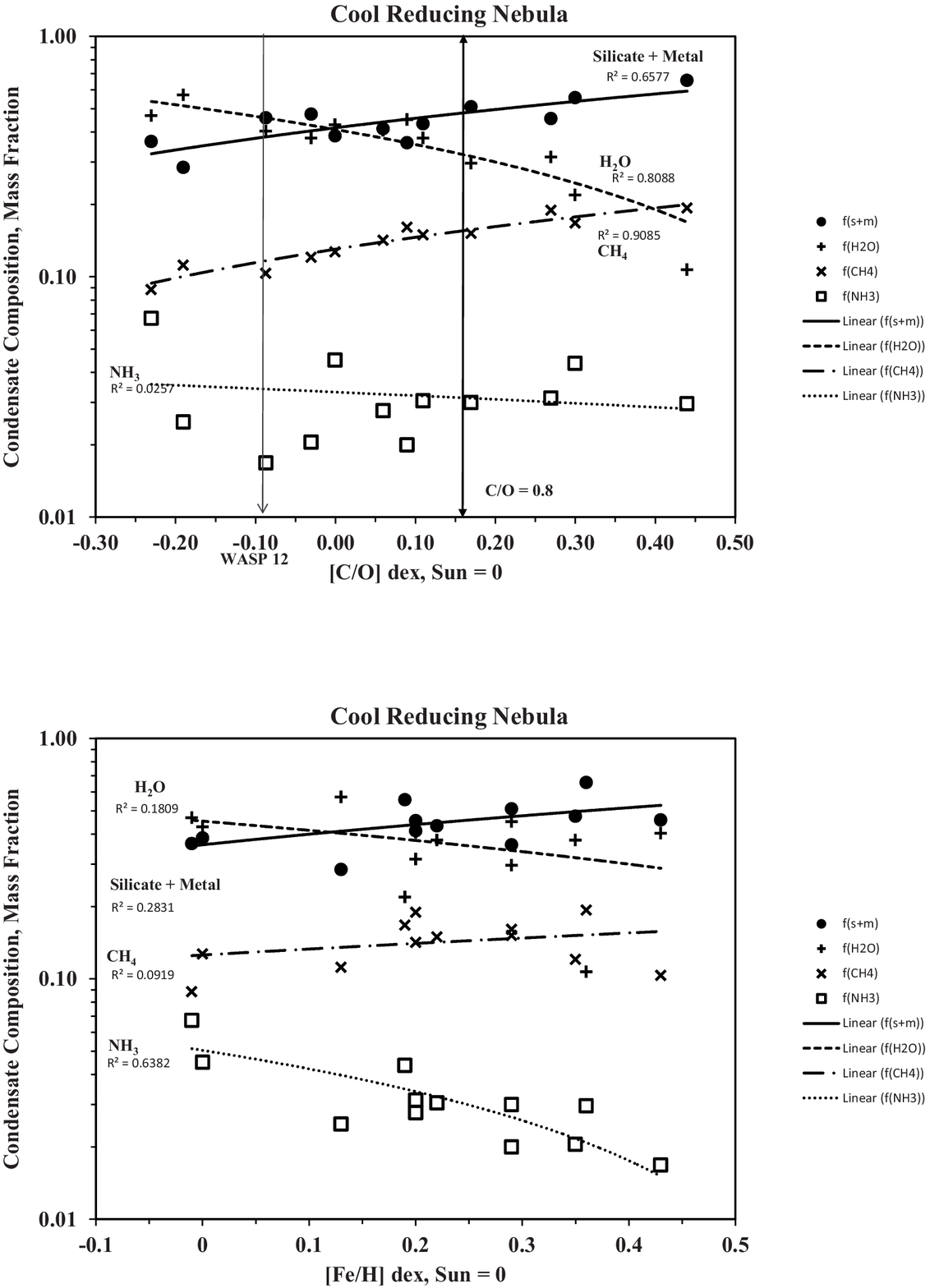}
\caption{Same as Fig. \ref{cool_oxy} but for reducing conditions.}
\label{cool_red}
\end{figure}

\clearpage
\begin{figure}
\centering
\includegraphics[angle=0,width=13cm]{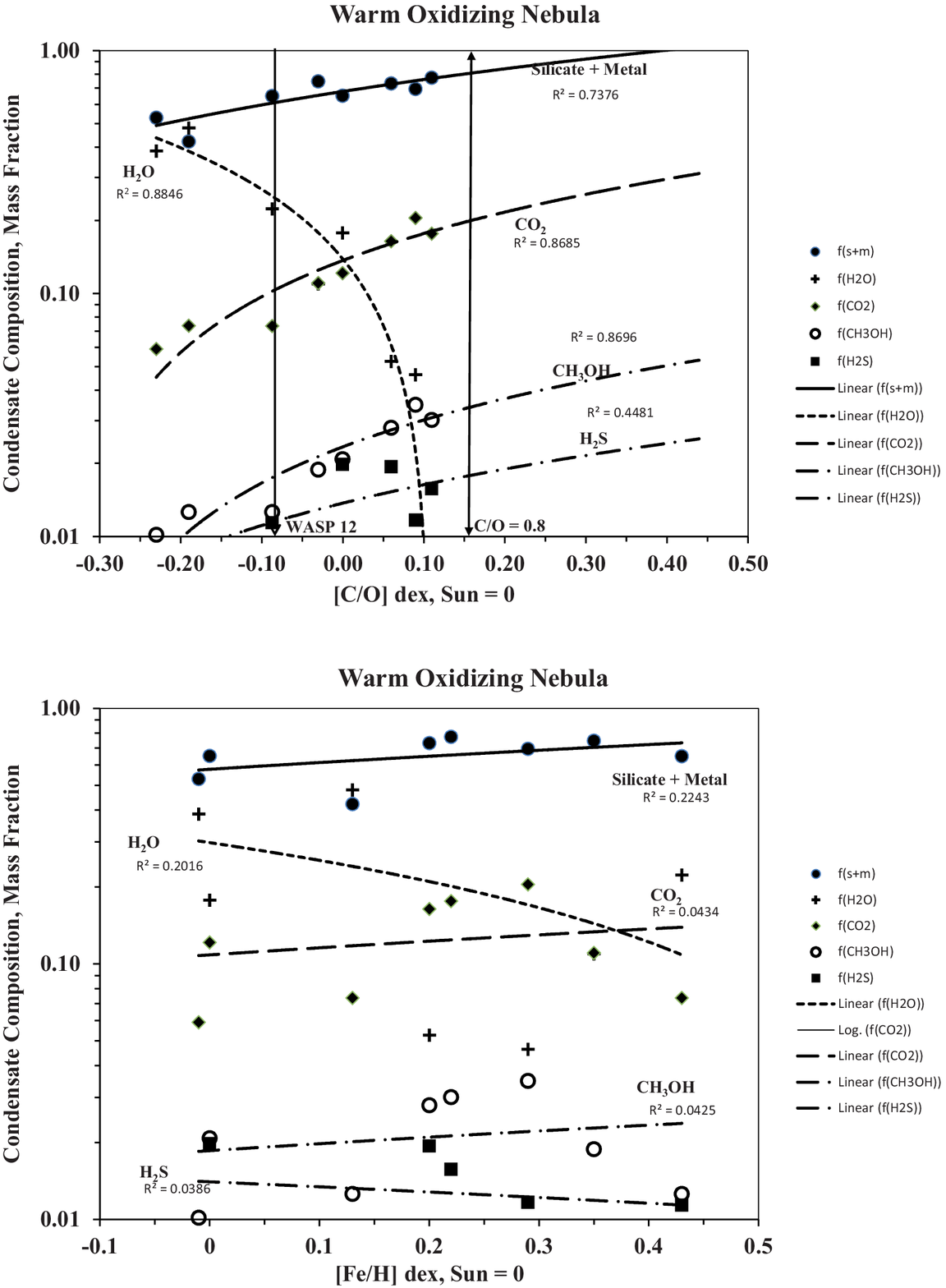}
\caption{Warm nebula case and oxidizing conditions.  Correlation of mass fractions of major ($>$ 0.01 on y-axis) refractory and volatile ice phases with [C/O] and [Fe/H].  Data are the same as Fig. \ref{comp2}a,b.  The correlation with [C/O] is much stronger than with [Fe/H] as indicated by the scatter in the calculated values and the R$^2$ values for the trendline analysis shown.}
\label{warm_oxy}
\end{figure}

\clearpage
\begin{figure}
\centering
\includegraphics[angle=0,width=13cm]{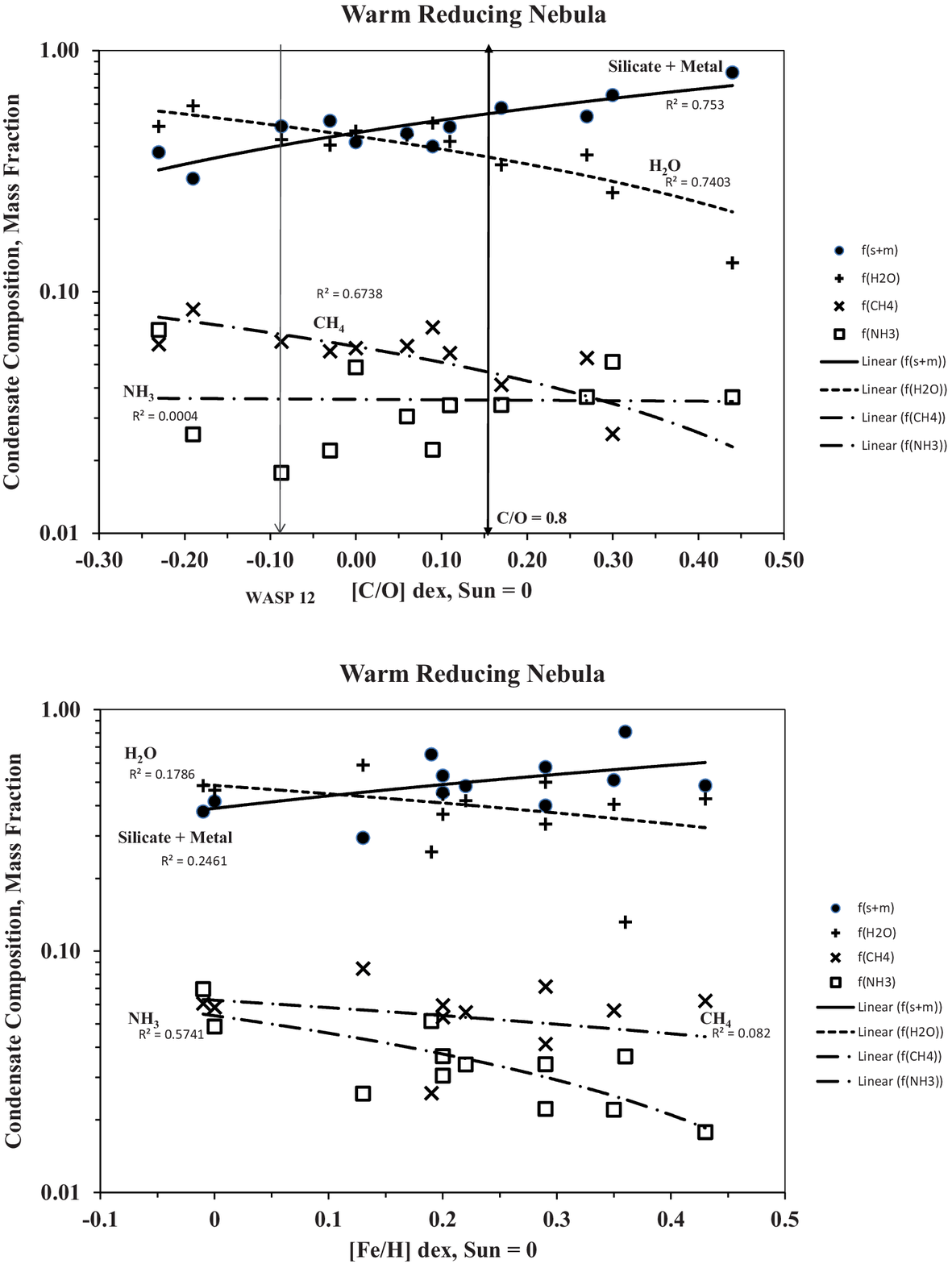}
\caption{Same as Fig \ref{warm_oxy} but for reducing conditions.}
\label{warm_red}
\end{figure}

\end{document}